\documentclass[runningheads,a4paper]{llncs}

\usepackage{amssymb}
\usepackage{graphicx}
\usepackage{caption}
\usepackage{subcaption}
\usepackage{amsmath}
\usepackage{subfig} 

\usepackage{url}

\newcommand{\DD}{\, \displaystyle}

\newcommand{\rond}[2]{\, \frac{\partial #1}{\partial #2}}
\newcommand{\fracc}[2]{\, \displaystyle \frac{ #1}{ #2}}

\newcommand{\morabba}[1]{\,\begin{flushright}
 \Rectsteel \\
\end{flushright}}

\newcommand{\CC}[2]{\, \binom{#1}{#2} 
}

\newcommand{\all}[2]{\,\begin{align}
                   #1 
                    \label{#2}
                   \end{align}
}

\begin{document}

\mainmatter  

\title{Beyond the Steady-State: Analytical Study of Network Growth  at  Arbitrary Times, for Arbitrary Initial Conditions }

\titlerunning{ Network Growth:  Arbitrary Times  and Initial Conditions}

%
%
\author{Babak Fotouhi$^{1,2}$   }
\authorrunning{Babak Fotouhi   }

\institute{$^1$ Department of Electrical and Computer Engineering\\
McGill University, Montr\'eal,  Canada\\
$^2$ Department of Sociology, 
McGill University, Montr\'eal,   Canada\\
\email{babak.fotouhi@mail.mcgill.ca} 
 }


%
%

\maketitle

\begin{abstract}
In studying network growth, the conventional approach is to devise a growth mechanism, quantify the evolution of a   statistic or distribution  (such as the degree distribution), and then solve the equations in the steady state (the infinite-size limit). Consequently, empirical studies also seek to verify the steady-state prediction in real data. The caveat concomitant with confining the analysis to this time regime is that  no real system has infinite size; most real growing  networks are far from the steady state. This underlines the importance  of finite-size analysis. In this paper, we  consider the shifted-linear   preferential   attachment  as an illustrative example of arbitrary-time network growth analysis. We  obtain the   degree distribution for arbitrary initial conditions  at  arbitrary times.  We corroborate our theoretical predictions with Monte Carlo simulations. 
\end{abstract}
 
\section{Introduction}

The network growth literature was commenced  by  the seminal   Barab{\'a}si-Albert model   posited initially  in~\cite{BA_1}. 
The motivation of studying the evolution of networks is to grasp underlying mechanisms that drive the growth process, which ramify into observable macro properties (such as a power-law degree distribution in the case of the  Barab{\'a}si-Albert model). 
   Examples of other growth models  include models with  edge growth~\cite{barab_link_growth_directed,redner3}, aging effects~\cite{klemm_1,age_dorog_1}, node deletion~\cite{newman_del_1,sarshar}, 
accelerated growth~\cite{dorog_accel_1},
   copying~\cite{kumar_copy,redner2}, and fitness-based models~\cite{Bian,fitness_1,pure_fit}.
These models are all purely structural, that is, the connectivity of new nodes is determined by factors that depend on time or structural measures of the network. 

In all the examples mentioned above, the prototypical way of analyzing a network growth model is as follows: a micro growth mechanism is devised (for example for the  Barab{\'a}si-Albert model, nodes enter sequentially and attach to existing nodes with degree-proportional probabilities), then the evolution of a desired quantity, such as the degree distribution is quantified (e.g., via  a rate equation), then the equations are simplified in the long time limit (the limit as $t \rightarrow \infty$), and then they are solved to yield steady-state quantities of interest. The practical shortcoming of such an approach is that no real network has infinite size, and the assumption of equilibrium is unrealistic for most networks. Note that this can lead to crucial  methodological consequences  concerning  the falsifiability of the model. For example, consider the  Barab{\'a}si-Albert model. It predicts that in the steady state, the network will exhibit a power law. In this case, if we take a real network and show that its degree distribution does not follow a power law, this will not refute the theory because one could counter with the contention that the network (no network, in fact) has infinite size, so the network at hand does not have the necessary condition (i.e., infinite size) for the prediction that theory provides. 

To alleviate this shortcoming and to remedy the falsifiability caveat, we proposed an alternative approach: focusing on the temporal evolution of the network for a given growth model. That is, instead of solving for  the quantities of interest in the steady-state, solve  for them   in  arbitrary times and for initial conditions. This way, for a given real network, an arbitrary point in time can be considered the origin of time and the network at that instant will constitute the initial condition. Then observations on the temporal evolution of desired quantities can be compared to the theoretical prediction, in order to assess the theory. 

In the present paper, we focus on a  shifted-linear preferential growth mechanism. We find the expected degree distribution as a function of time, for arbitrary initial conditions and arbitrary times. The results are corroborated with Monte Carlo simulations throughout the paper.

\section{Model}
The growth process starts from a given initial network with $N(0)$ nodes and $L(0)$ links, with known degree distribution $p_k(0)$. The network grows via the successive addition of new nodes. At each time step  a  new node is born, and it forms $\beta$ links to existing nodes, according to the  preferential linking: the probability that an existing node $x$ receives a  link  from the new node at time $t$ is proportional to $k_x(t)+\theta$, where $k_x(t)$ is the degree of node $x$ at time $t$, and $\theta$ is the initial attractiveness, a  positive constant of the model. To obtain normalized probabilities, we need to divide $k_x(t)+\theta$ for each $x$ by the sum of this quantity over every node. The sum over $k_x$ yields twice the number of links at time $t$, which is $2L(0)+2\beta t$. The sum over $\theta$ yields $\theta N(t)$, where $N(t)$ is the number of nodes at time $t$, which equals $N(0)+t$. Thus, the probability that node $x$ receives a link emanated from the newly-born node equals 
\all{
\pi_x(t)=\frac{k_x+\theta}{2L(0)+2\beta t+\theta N(0) + \theta t}
.}{pi_x_0}
Hereinafter, we will denote $2L(0)+\theta N(0)$ by $\lambda$, and $2\beta+\theta$ by $\nu$. So~\eqref{pi_x_0} transforms into
\all{
\pi_x(t)=\frac{k_x+\theta}{\lambda + \nu t}
.}{pi_x}


\section{Evolution of the Degrees}
At each timestep, we can quantify the expected the change in $N_k(t)$, which is the number of nodes in the network that have degree $k$ at time $t$.  The value of $N_k(t)$ can be altered if   at time $t$,  an existing  node with degree $k$ receives a link   from the newly-born node (which would increment the degree of the receiving node to $k+1$, decrementing $N_k$), or if  an existing node with degree $k-1$  receives a link (which would increment the degree of the receiving node to $k$, incrementing $N_k$). For each incoming node, $N_\beta (t)$ increments. The following rate equation quantifies the evolution of $N_k(t)$: 

\all{
N_k(t+1)-N_k(t)&= \beta   \left[\fracc{(k-1+\theta)N_{k-1}(t)-(k+\theta)N_k(t)}{\lambda + \nu t}   \right]
+   \delta_{k,\beta}.
}{Nk_dot_0}

This can be rearranged and expressed equivalently as follows

\all{
N_k(t+1)-N_k(t)&= \beta      \left[\fracc{(k-1+\theta)N_{k-1}(t)-(k+\theta)N_k(t)}{\lambda + \nu t}   \right]
+  \delta_{k,\beta}.
}{Nk_dot_00}

This is a two-dimensional difference equation in time and $k$. We now employ a time-continuous approximation to this equation and replace it with the following difference-differential equation:

\all{
\rond{N_k(t)}{t}&= \beta      \left[\fracc{(k-1+\theta)N_{k-1}(t)-(k+\theta)N_k(t)}{\lambda + \nu t}   \right]
+   \delta_{k,\beta}.
}{Nk_dot}

Note that the relative  error of this approximation at each timestep is proportional to  $\frac{1}{\lambda + \nu t}$. Even for short times, the $2L(0)$ in the denominator ensures that the truncation error is small, provided that $L(0)$ is large. In our Monte Carlo simulations, the initial network comprises 100 links, and the predictions are remarkably accurate. Note that 100 links is tiny as compared to many real networks, such as social and biological networks, citation and collaboration networks, the web and other online networks. The typical size of these networks are way larger than 100, so the approximation will be   conservative in real settings. In Monte Carlo simulations, we also tested larger initial networks and verified the accuracy of predictions for large systems. Simulation results  are  presented in Section~\ref{sec:simul}. Figure~\ref{figs} illustrates the remarkable accuracy of the theoretical predictions, where the error bars are smaller than the markers used for depiction. 

To solve~\eqref{Nk_dot}, we define the generating function: 
\all{
\psi(z,t) \stackrel{\text{def}}{=} \DD \sum_{k=1}^{\infty} N_k(t) z^{-k}
.}{psi_def}

This is the conventional Z transform. We multiply both sides of~\eqref{Nk_dot} by $z^{-k}$ and sum over $k$. The left hand side yields $\rond{\psi}{t}$. For the  terms on the right hand side, we use two standard properties of the Z-transform:  if the  generation function  of some sequence $a_k$ is given by $A(z)$, then  (1) the generating function for sequence $ka_k$ is given by $-z \frac{d A(z)}{dz}$, and  (2) the generating function for the sequence $a_{k-1}$ is given by $z^{-1}a_k$. Using these two properties, Equation~\eqref{Nk_dot} yields
\all{
\DD \rond{\psi(z,t)}{t} = \fracc{\beta     }{\lambda + \nu t}  \Bigg[ (z-1) \rond{\psi(z,t)}{z} + \theta (z^{-1}-1) \psi(z,t) \Bigg] +   z^{-\beta}
.}{psi_dot_1}
This can be rearranged and recast as
\all{
\resizebox{.95 \linewidth}{!}{$
\DD \rond{\psi(z,t)}{t} - \fracc{\beta     }{\lambda + \nu t}    (z-1) \rond{\psi(z,t)}{z} =  \fracc{\beta     \theta}{\lambda + \nu t} (z^{-1}-1) \psi(z,t)    +   z^{-\beta}
$}
.}{psi_dot_1}
In Appendix~\ref{app:pde} we solve this partial differential equation. Let us define
\all{
\begin{cases}
c  \stackrel{\text{def}}{=} 1 - \left( \fracc {\lambda}{\lambda+\nu t}\right)^{\frac{\beta     }{\nu}} \\ \\
F(z,t) \stackrel{\text{def}}{=} \fracc{    (\lambda+\nu t)  }{\beta     }  \DD  \sum_{m=0}^{\infty}  (-1)^{m}\fracc{ z^{\frac{-\nu}{\beta     }-m-\beta}}{\frac{\nu}{\beta     } +m+\beta+\theta}
.
\end{cases}
 }{defs}
Using these definitions,  the solution to~\eqref{psi_dot_1} reads
 \all{
 \psi(z,t) 
 = F(z,t) 
 +  z^{\theta}
  \left(\fracc{z-c}{1-c}\right)^{-\theta}
 \bigg[\psi \left( \fracc{z-c}{1-c} , 0\right) -F \left(\fracc{z-c}{1-c},0 \right) \bigg]
 .}{psi_sol}
Note that $\psi \left( \frac{z-c}{1-c} , 0\right)$ is obtained by taking the Z-transform of the sequence $N_k(0)$ (which is given as the initial condition) and then replacing $z$ by $\frac{z-c}{1-c}$.  In Appendix~\ref{app:inv} we take the inverse transform of this expression to obtain $N_k(t)$. Let us define 
\all{
c  \stackrel{\text{def}}{=} 1 - \left( \fracc {\lambda}{\lambda+\nu t}\right)^{\frac{\beta     }{\nu}}
.}{c_def_main}
Using this definition for brevity, the inverse transform of~\eqref{psi_sol} reads

\all{
&N_k(t)
= 
(1-c)^{\theta} c^k  \DD \sum_{r=1} ^{k} N_k(0)  \left( \fracc{1-c}{c} \right)^r \CC{k+\theta-1}{r+\theta-1}
\nonumber \\ &
 + \DD  
 \fracc{  \bigg[\lambda+ (2\beta +\theta)   t\bigg]   }{       \beta  }  
\fracc{\Gamma(k+\theta) }{ \Gamma(\beta+\theta)}  \fracc{\Gamma \left( \beta+ 2+\frac{\theta}{\beta} +\theta  \right) }
{\Gamma \left( k+3+ \frac{\theta}{\beta}  + \theta \right) }  u(k-\beta)
\nonumber \\
&
\resizebox{.98 \linewidth}{!}{$
-  \fracc{   \lambda   (1-c)^{\theta} c^k  }{      \beta  }  
  \Gamma \left( \beta+  2+\frac{\theta}{\beta} +\theta  \right)
\fracc{\Gamma(k+\theta) }{ \Gamma(\beta+\theta)}
\DD \sum_{r=\beta} ^{k}
 \displaystyle \frac{ \left( \fracc{1-c}{c} \right)^r  }{(k-r)! \Gamma \left( m+3+ \frac{\theta}{\beta}  + \theta \right)}
$}.
}{Nk_ultimate}

We can divide this by the number of nodes at time $t$ to obtain the fraction of nodes with degree $k$ at time $t$. The result is

\all{
&P_k(t)
= 
(1-c)^{\theta} c^k  \fracc{N(0)}{N(0)+t}\DD \sum_{r=1} ^{k} P_k(0)  \left( \fracc{1-c}{c} \right)^r \CC{k+\theta-1}{r+\theta-1}
\nonumber \\ &
 + \DD  
 \fracc{ 1 }{       \beta  }   \fracc{ \lambda+ (2\beta +\theta)   t  }{N(0)+t}
\fracc{\Gamma(k+\theta) }{ \Gamma(\beta+\theta)} \fracc{\Gamma \left( \beta+  2+\frac{\theta}{\beta} +\theta  \right) }
{\Gamma \left( k+3+ +\frac{\theta}{\beta}  + \theta \right) }  u(k-\beta)
\nonumber \\
&
\resizebox{.98 \linewidth}{!}{$
-  \fracc{   \lambda   (1-c)^{\theta} c^k  }{      \beta  }  
\fracc{  \Gamma \left( \beta+  2+\frac{\theta}{\beta} +\theta  \right)}{N(0)+t}
\fracc{\Gamma(k+\theta) }{ \Gamma(\beta+\theta)}
\DD \sum_{r=\beta} ^{k}
 \displaystyle \frac{ \left( \fracc{1-c}{c} \right)^r  }{(k-r)! \Gamma \left( m+3+ \frac{\theta}{\beta}  + \theta \right)}
$}.
}{Pk_ultimate}

The first term is the effect of initial nodes. In the long time limit, the $N(0)+t$ in the denominator makes this term vanish. Moreover, from~\eqref{c_def_main} we observe that in the limit as $t \rightarrow \infty$, we have $c \rightarrow 1$. This means that every $(1-c)^r$ term as well as the $(1-c)^{\theta}$ prefactor all tend to zero in the long time limit. Note that the $c^r$ in the denominator will not cause divergence, because the $c^k$ prefactor removes the singularity. So the first term on the right hand side of~\eqref{Pk_ultimate} vanishes in the long time limit, as we intuitively expect. 

The second term on the right hand side of~\eqref{Pk_ultimate} reaches a horizontal asymptote in the long time limit. In this limit, we have $   \frac{ \lambda+ (2\beta +\theta)   t  }{N(0)+t} \rightarrow 2\beta+\theta$. 

Finally, the last term on the right hand side of~\eqref{Pk_ultimate} vanishes in the long time limit for the same reasons delineated above for the first term. So in the steady state, the first and third terms have no share in the degree distribution, and the second term dominates. We have: 

\all{
\lim_{t\rightarrow \infty} P_k(t)= 
 \DD  
\left( 2+\fracc{\theta}{\beta}\right)
\fracc{\Gamma(k+\theta) }{ \Gamma(\beta+\theta)} \fracc{\Gamma \left( \beta+  2+\frac{\theta}{\beta} +\theta  \right) }
{\Gamma \left( k+3+\frac{\theta}{\beta} + \theta \right) }  u(k-\beta)
.}{Pk_ss}

%
%

This is in agreement with the results in~\cite{dorog_nk}. 
Finally, we can set $\theta=0$ to recover the degree distribution of the conventional  Barab{\'a}si-Albert model: 

\all{
\lim_{t\rightarrow \infty} P_k^{\textnormal{BA}}(t) = 
 \DD  
 2
\fracc{(k-1)! }{(\beta-1)!} \fracc{(\beta+1)! }
{(k+2)! }  u(k-\beta) = \fracc{2 \beta(\beta+1)}{k(k+1)(k+2)} u(k-\beta)
.}{Pk_ss}

This is in agreement with the long-known result, as given for example in~\cite{dorog_nk,redner1,Bol}.

\section{Simulation Results}\label{sec:simul}

\begin{figure}[!Ht]
        \centering
      \begin{subfigure}[b]{.5 \textwidth}
              \includegraphics[width=\textwidth,height=5cm]{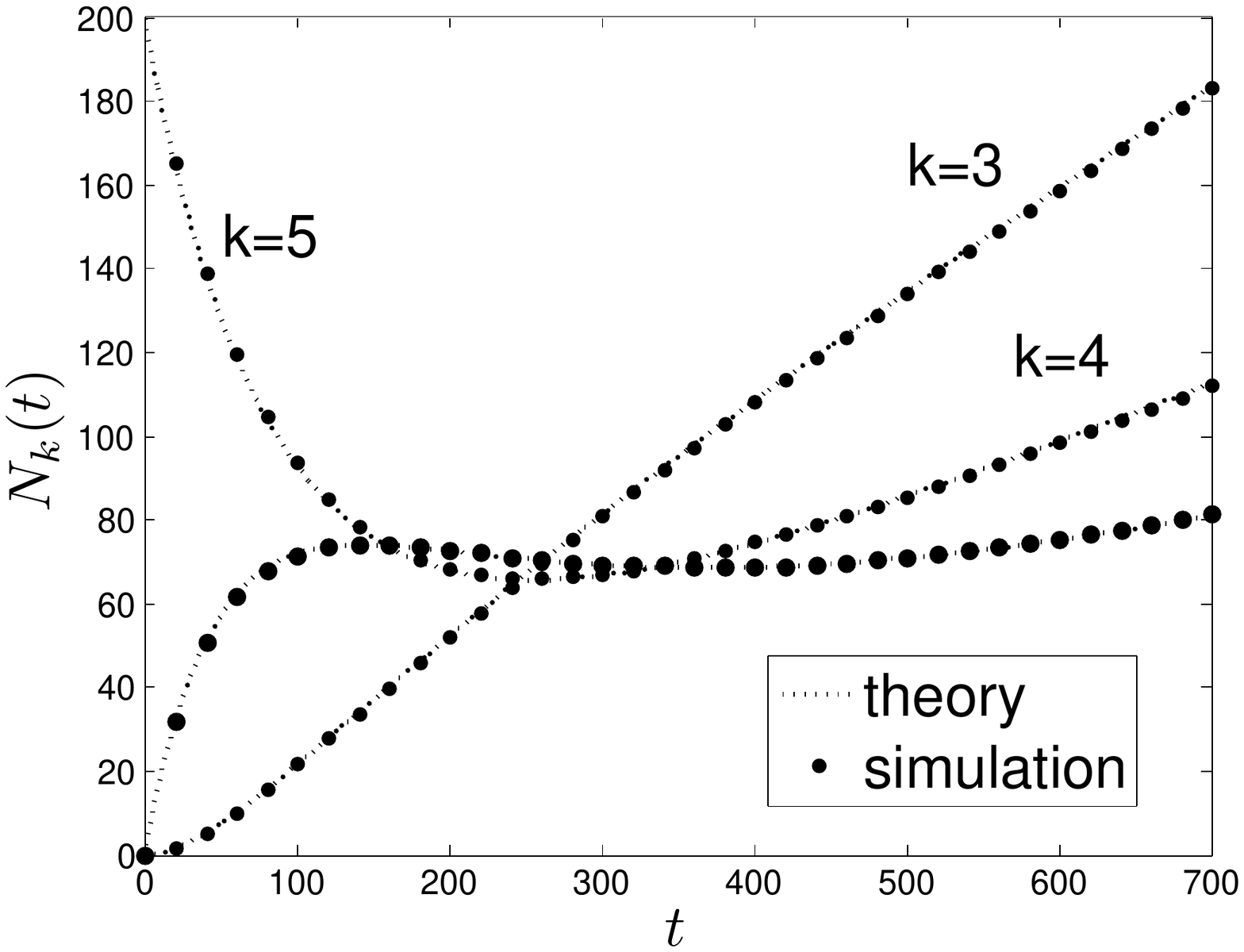}
                \caption{
  $\beta=2$, $\theta=20$, $N(0)=200$. 
}
                \label{fig_1}
        \end{subfigure}%
        \begin{subfigure}[b]{0.5\textwidth}
                \includegraphics[width=\textwidth,height=5cm]{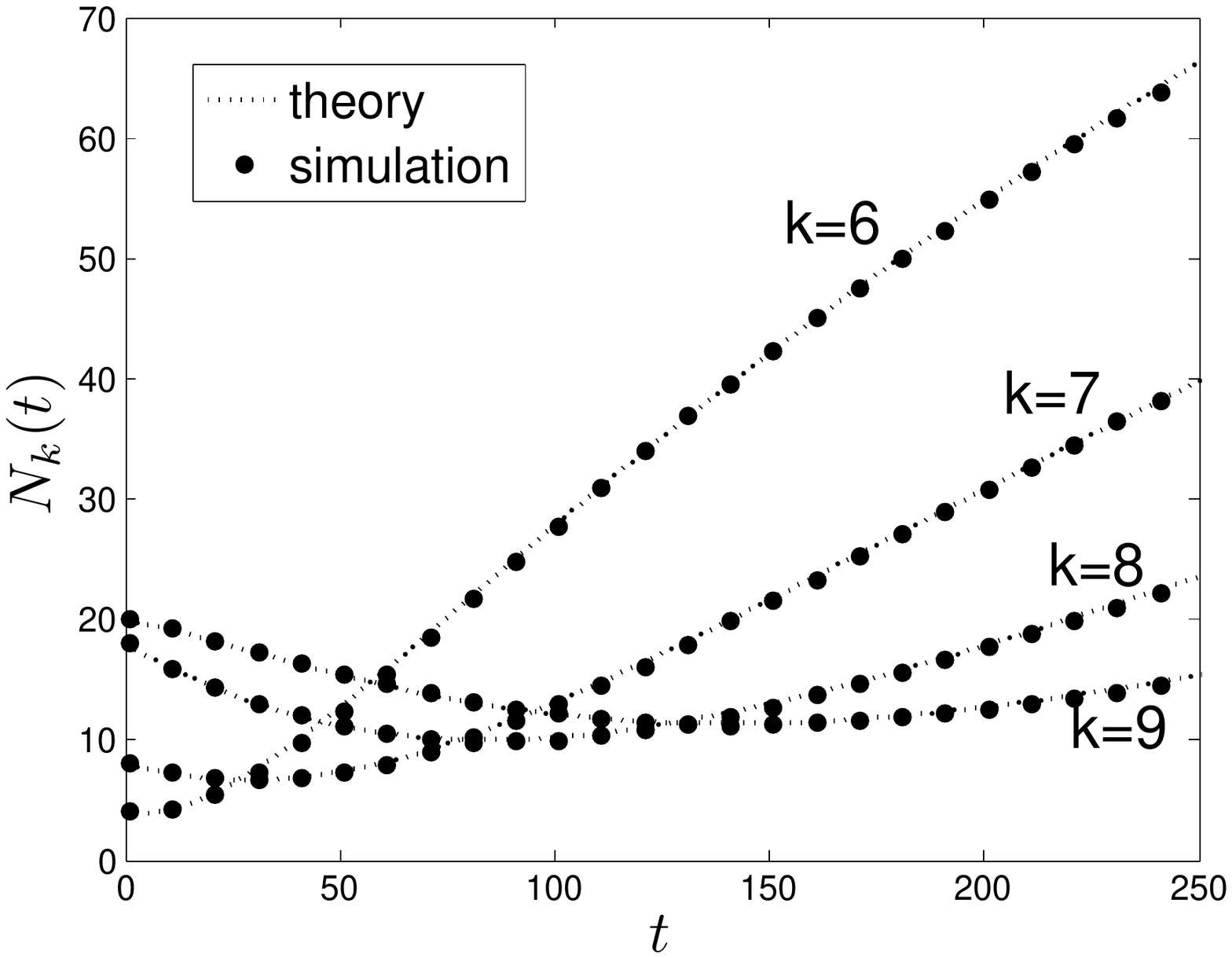}
                \caption{
           $\beta=5$, $\theta=4.4$, $N(0)=200$.
                 }
                \label{fig_2}
        \end{subfigure}%
\\
 \begin{subfigure}[b]{0.5\textwidth}
                \includegraphics[width=\textwidth,height=5cm]{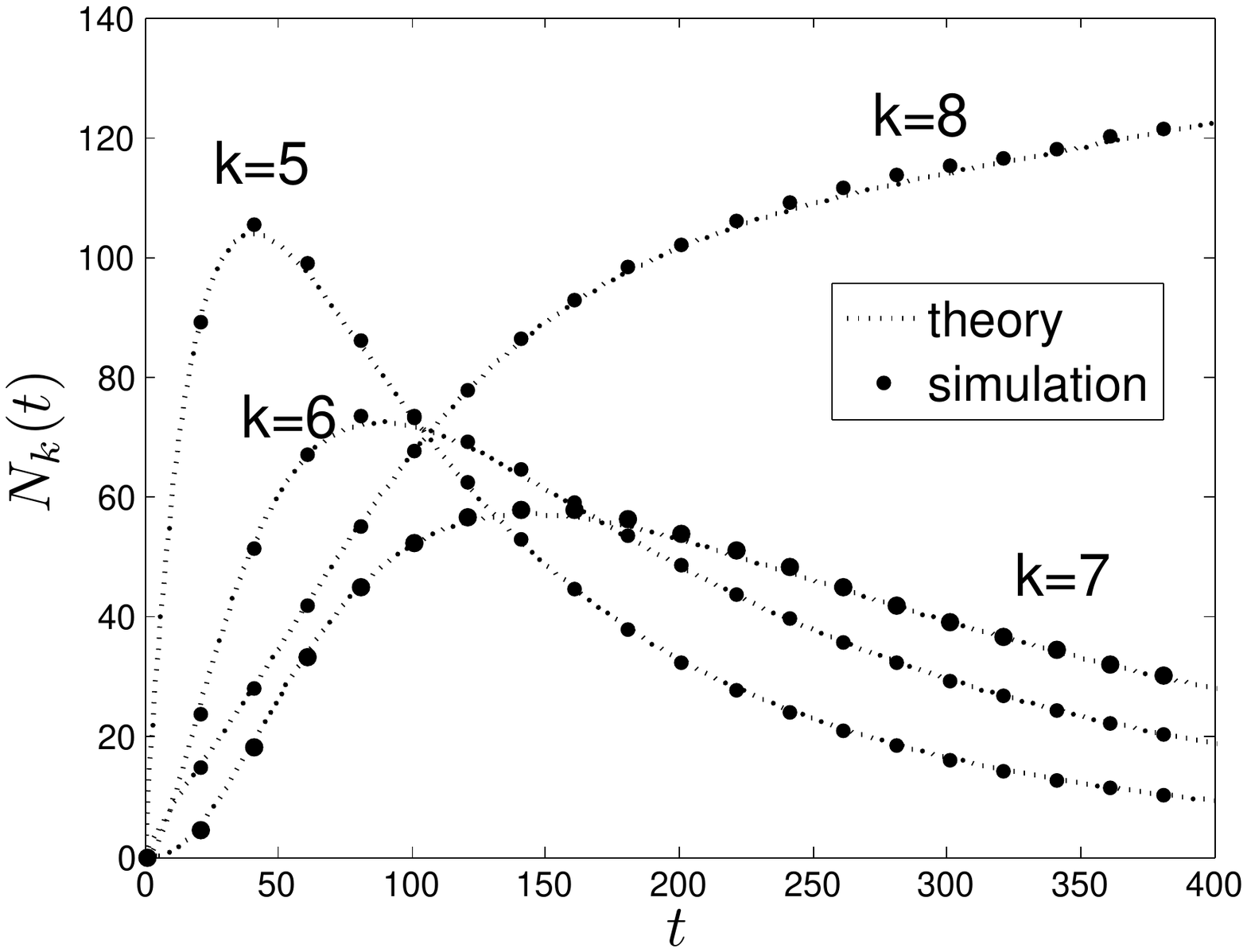}
                \caption{
       $\beta=8$, $\theta=4$, $N(0)=300$.
                 }
                \label{fig_3}
        \end{subfigure}%
        \begin{subfigure}[b]{0.5\textwidth}
                \includegraphics[width= \textwidth,height=5cm]{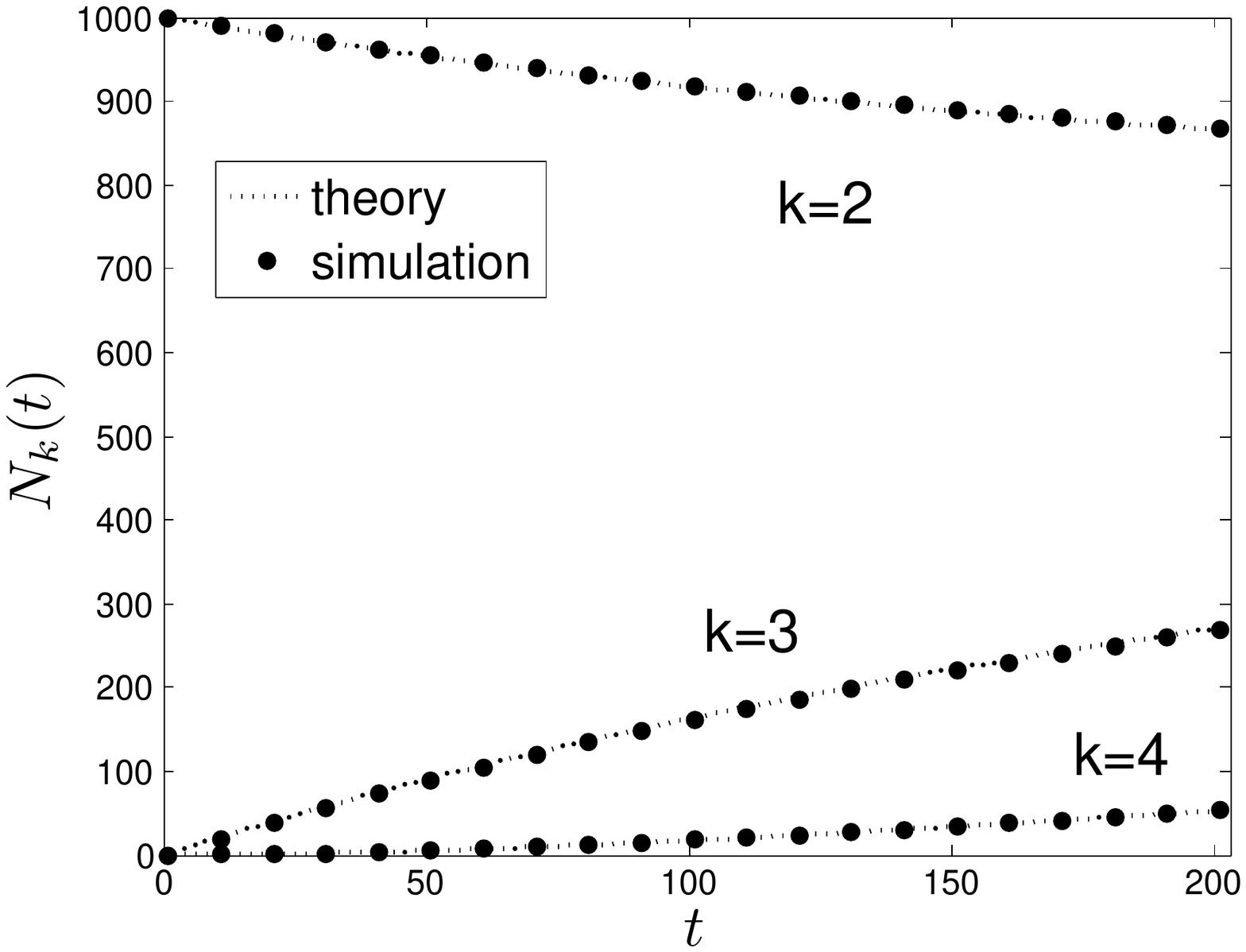}
                \caption{
 $\beta=2$, $\theta=4.4$, $N(0)=1000$.
               }
                \label{fig_4}
        \end{subfigure}
         \caption{
         \footnotesize{Simulation results and theoretical predictions of Equation~\eqref{Nk_ultimate}. The number of Monte Carlo trials for all  settings is 1000. The initial network in Figures (a) and (c) are 4-regular rings, that of Figure (b) is a small-world that is constructed by taking a ring and establishing every non-existing node with probability 0.05, and the initial network of Figure (d) is a ring. }
          }
\label{figs}
\end{figure}

Figure~\ref{fig_1} shows the simulation results for the a 4-regular ring of 200 nodes as the initial network (a 4-regular ring is obtained by connecting every second neighbor on a ring). The temporal evolution of $N_k(t)$ for three distinct values of $k$ are depicted. As can be seen, theoretical predictions match the simulation results to high accuracy. The error bars are of the size of the markers used in the graph. The value of $\beta$ is 2 and $\theta$ is 20. Since $\theta$ is large, the preferential share of the growth kernel is weaker, that is. Generally, the growth process  is closer to a uniform growth as $\theta$ increases.  

Figure~\ref{fig_2} depicts the simulation results and theoretical predictions for a small-world network, which is constructed by taking a ring of 200 nodes and establishing every non-existing link with probability 0.05.

The initial network in Figure~\ref{fig_3} a 4-regular ring of 300 nodes. Figure~\ref{fig_4} presents the simulation results for a larger initial network: a ring of 1000 nodes. It can be observed that even for short times, the accuracy of  the theoretical predictions are remarkably high.

\section{Summary and Discussion}

We contended that conventional emphasis on the steady state for the analysis of network growth models   might  engender methodological caveats. We proposed focusing on arbitrary times and solving the growth problem for arbitrary initial conditions. We considered the shifted-linear growth scheme and obtained the degree distribution as a function of time for arbitrary initial conditions. 
 We verified the theoretical predictions with Monte Carlo simulations. 
 
 Plausible extensions to this result include moving beyond the expected degree distribution, and obtaining the distribution of degree distributions. This is essential for devising rigorous statistical tests to assess network growth mechanisms upon observing  longitudinal  data  on  the  degree distribution. For a given initial network, there are various paths that the system can take, due to the random nature of the growth process. What we obtained in this paper is the average over all those paths. But at each timestep, there is a distribution over all those paths. If we knew that distribution, then we could  devise statistical recipes for hypothesis testing, to be able to answer questions of the form: ``if the posited growth mechanism is true, what is the probability that the empirical degree distribution at time $t$ would occur?". 
  Since a hypothesis testing scheme like the one discussed requires  a null hypothesis, it would be plausible to solve the same problem for the uniformly-growing networks. So the problem is:  under uniform growth,  for a given initial network, what is the probability that at time $t$, the number of nodes with degree $k$ would be  $N_k$?

Finally, it would be plausible to consider time--reversal in network growth problems. Given a growth mechanism and the degree distribution at time $t$, what is the  distribution   that the system has most likely been at time $t-s$?  What information does  observing the network at some state at time $t$ (say, observing its degree distribution at time $t$) give about the   states (degree distributions) of the system at earlier times?

\appendix
\section{Solving the PDE in~\eqref{psi_dot_1}}\label{app:pde}
The PDE we need to solve is:
\all{
\resizebox{.95 \linewidth}{!}{$
\DD \rond{\psi(z,t)}{t} - \fracc{\beta     }{\lambda + \nu t}    (z-1) \rond{\psi(z,t)}{z} =  \fracc{\beta     \theta}{\lambda + \nu t} (z^{-1}-1) \psi(z,t)    +   z^{-\beta}
$}
.}{psi_dot_1_app}
We  employ the method of characteristic curves to solve this equation (see for example~\cite{PDE}, for   background on this method). We need to first solve the following system of equations:
\all{
\fracc{dt}{1}= \fracc{dz}{- \fracc{\beta     }{\lambda + \nu t}    (z-1)}
= \fracc{d \psi}{\fracc{\beta     \theta}{\lambda + \nu t} (z^{-1}-1) \psi(z,t)    +   z^{-\beta}}
.}{sys}

From the first equation we get
\all{
\fracc{dt}{1}= \fracc{dz}{- \fracc{\beta     }{\lambda + \nu t}    (z-1)}
\Longrightarrow 
(z-1)^{\frac{\nu}{\beta     }} (\lambda + \nu t) = C
.}{sys1}

The second equation is
\all{
 \fracc{dz}{- \fracc{\beta     }{\lambda + \nu t}    (z-1)}
= \fracc{d \psi}{\fracc{\beta     \theta}{\lambda + \nu t} (z^{-1}-1) \psi(z,t)    +   z^{-\beta}}
.}{sys2}

This can be rearranged and rewritten as follows
 
\all{
\fracc{d\psi}{dz}- \frac{\theta}{z} \psi = \fracc{- z^{-\beta} (\lambda+\nu t)}{\beta     (z-1)}
.}{sys2_3}

Using~\eqref{sys1}, this transforms into

\all{
\fracc{d\psi}{dz}- \frac{\theta}{z} \psi = \fracc{-  C z^{-\beta}  }{\beta     }(z-1)^{\frac{-\nu}{\beta     }-1}
.}{sys2_3}

This is an ordinary  first-order linear differential equation, with integrating factor $z^{-\theta}$. The solution is given by
\all{
\psi=z^{\theta} \left[ \fracc{-   C }{\beta     }\DD \int^z z^{-\beta-\theta} (z-1)^{\frac{-\nu}{\beta     }-1}  dz + \Phi(C) \right] 
,}{psi_sol_1}

where $\Phi(C)$, according to the method of characteristics, is an arbitrary function of $C$ that is uniquely specified for given initial conditions. We expand the integrand before performing the integration. We have
\all{
 (z-1)^{\frac{-\nu}{\beta     }-1} =\DD  z^{\frac{-\nu}{\beta     }-1} \DD \sum_{m=0}^{\infty}  (-1)^m \binom{\frac{ \nu}{\beta     }+m }{m} z^{-m}
.}{tylor_1}
 Plugging this into~\eqref{psi_sol_1}, we get
 \all{
 \psi(z,t)&=z^{\theta} \left[ \fracc{-     C  }{\beta     }  \DD  \sum_{m=0}^{\infty}  (-1)^m \int^z  z^{\frac{-\nu}{\beta     }-1-m-\beta-\theta}
 +\Phi(C) \right]
 \nonumber \\ 
 &
 =z^{\theta} \left[ \fracc{+   C }{\beta     }  \DD  \sum_{m=0}^{\infty}  (-1)^{m}\fracc{ z^{\frac{-\nu}{\beta     }-m-\beta-\theta}}{\frac{\nu}{\beta     } +m+\beta+\theta}
 +\Phi(C) \right]
  \nonumber \\ 
 &
 =  \fracc{    C  }{\beta     }  \DD  \sum_{m=0}^{\infty}  (-1)^{m}\fracc{ z^{\frac{-\nu}{\beta     }-m-\beta}}{\frac{\nu}{\beta     } +m+\beta+\theta}
 +\Phi(C)z^{\theta}
 .}{psi_7}
Now we use~\eqref{sys1} to plug in the explicit expression for $C$ into~\eqref{psi_7}:
\all{
 (z-1)^{\frac{-\nu}{\beta     }-1} =\DD  z^{\frac{-\nu}{\beta     }-1} \DD \sum_{m=0}^{\infty}  (-1)^m \binom{\frac{ \nu}{\beta     }+m }{m} z^{-m}
.}{tylor_1}
 Plugging this into~\eqref{psi_sol_1}, we get
 \all{
 \psi(z,t) 
 =  &\fracc{    C }{\beta     }  \DD  \sum_{m=0}^{\infty}  (-1)^{m}\fracc{ z^{\frac{-\nu}{\beta     }-m-\beta}}{\frac{\nu}{\beta     } +m+\beta+\theta}
 \nonumber \\ &
 +\Phi \left[ (z-1)^{\frac{\nu}{\beta     }} (\lambda + \nu t)  \right] z^{\theta}
 .}{psi_8}

Let us define
\all{
F(z ) \stackrel{\text{def}}{=} \fracc{    (z-1)^{\frac{\nu}{\beta     }}    }{\beta     }  \DD  \sum_{m=0}^{\infty}  (-1)^{m}  \binom{\frac{ \nu}{\beta     }+m }{m}\fracc{ z^{\frac{-\nu}{\beta     }-m-\beta}}{\frac{\nu}{\beta     } +m+\beta+\theta}
}{F_def}

Then~\eqref{psi_8} can be rewritten as follows: 
 \all{
 \psi(z,t) 
 =  (\lambda+\nu t) F(z ) 
 + z^{\theta}\Phi \left[ (z-1)^{\frac{\nu}{\beta     }} (\lambda + \nu t)  \right]
 .}{psi_9}

We need to uniquely determine $\Phi(\cdot)$. At time $t=0$, Equation~\eqref{psi_9} becomes
 \all{
& \psi(z,0) 
 =\lambda  F(z ) 
 + z^{\theta} \Phi \left[ (z-1)^{\frac{\nu}{\beta     }}  \lambda    \right]\Longrightarrow \nonumber \\  &
 \Phi \left[ (z-1)^{\frac{\nu}{\beta     }}  \lambda    \right] =z^{-\theta} \bigg[\psi(z,0)- \lambda F(z )\bigg]
 \Longrightarrow   \nonumber \\&
 \Phi(X)=\left[ \left(\frac{X}{\lambda}\right)^{\frac{\beta     }{\nu}}+1\right]^{-\theta}
 \bigg[\psi \left(  \left(\frac{X}{\lambda}\right)^{\frac{\beta     }{\nu}}+1   , 0\right) - \lambda F \left( \left(\frac{X}{\lambda}\right)^{\frac{\beta     }{\nu}}+1  \right) \bigg]
 .
 }{psi_10}
Also, let us define
\all{
c  \stackrel{\text{def}}{=} 1 - \left( \fracc {\lambda}{\lambda+\nu t}\right)^{\frac{\beta     }{\nu}}
.}{c_def}
Then it follows that
\all{
(z-1) \left(\frac{\lambda+\nu t}{\lambda}\right)^{\frac{\beta     }{\nu}}+1=\fracc{z-c}{1-c}
.}{jaleb}
Using~\eqref{jaleb},~\eqref{c_def}, we can simplify~\eqref{psi_10} into the following:
\all{
z^{\theta}  \Phi \left[ (z-1)^{\frac{\nu}{\beta     }} (\lambda + \nu t)  \right] 
  = 
  z^{\theta}
  \left(\fracc{z-c}{1-c}\right)^{-\theta}
 \bigg[\psi \left( \fracc{z-c}{1-c} , 0\right) - \lambda F \left(\fracc{z-c}{1-c}  \right) \bigg]
 }{phi_final}
Substituting the last term on the right hand side of~\eqref{psi_9} with the expression in~\eqref{phi_final}, we arrive at
 \all{
 \psi(z,t) 
 =  (\lambda+\nu t) F(z) 
 +  z^{\theta}
  \left(\fracc{z-c}{1-c}\right)^{-\theta}
 \bigg[\psi \left( \fracc{z-c}{1-c} , 0\right) -\lambda F \left(\fracc{z-c}{1-c}  \right) \bigg]
 .}{psi_9}

\section{Taking the Inverse Transform of~\eqref{psi_sol}} \label{app:inv}
Now we need to take the inverse transform of this expression. We do this term by term. First, we take the inverse transform of $F(z)$. We have

\all{
F(z ) &= \fracc{   (z-1)^{\frac{\nu}{\beta     }}    }{\beta     }  \DD  \sum_{m=0}^{\infty}  (-1)^{m} \binom{\frac{ \nu}{\beta     }+m }{m} \fracc{ z^{\frac{-\nu}{\beta     }-m-\beta}}{\frac{\nu}{\beta     } +m+\beta+\theta}
\nonumber \\ & 
=  \fracc{  z^{\frac{\nu}{\beta     }}   (1-z^{-1})^{\frac{\nu}{\beta     }} }{\beta     }  
  \DD  \sum_{m=0}^{\infty}  (-1)^{m} \binom{\frac{ \nu}{\beta     }+m }{m} \fracc{ z^{\frac{-\nu}{\beta     }-m-\beta}}{\frac{\nu}{\beta     } +m+\beta+\theta}
  \nonumber \\ &
  =
   \fracc{      (1-z^{-1})^{\frac{\nu}{\beta     }} }{\beta     }  
  \DD  \sum_{m=0}^{\infty}  (-1)^{m} \binom{\frac{ \nu}{\beta     }+m }{m} \fracc{ z^{-m-\beta}}{\frac{\nu}{\beta     } +m+\beta+\theta}
    \nonumber \\ &
  =
    \fracc{1    }{\beta     }  
  \DD  \sum_{m,r}  (-1)^{m+r}  \fracc{z^{-m-\beta-r}}{\frac{\nu}{\beta     } +m+\beta+\theta}
 \binom{\frac{ \nu}{\beta     }+m }{m}  \binom{\frac{ \nu}{\beta     }}{r}.
}{F_inv_0}

The inverse Z-transform of $z^{-a}$ for some integer $a$ is $\delta[k-a]$. So we take the inverse transform of $F(z)$ term by term:

\all{
F(z )   \xrightarrow{\mathcal{Z}^{-1} }
 &
    \fracc{1     }{\beta     }  
  \DD  \sum_{m,r}  (-1)^{m+r}  \fracc{\delta[k-m-\beta-r]}{\frac{\nu}{\beta     } +m+\beta+\theta}
 \binom{\frac{ \nu}{\beta     }+m }{m}  \binom{\frac{ \nu}{\beta     }}{r}
 \nonumber \\ &
    \fracc{1     }{\beta     }  
  \DD  \sum_{m}   \fracc{ (-1)^{k-\beta}}{\frac{\nu}{\beta     } +m+\beta+\theta}
 \binom{\frac{ \nu}{\beta     }+m }{m}  \binom{\frac{ \nu}{\beta     }}{k-m-\beta}
}{F_inv_1}

Now we utilize the following identity: 
\all{  
  \DD  \sum_{m}   \fracc{ (-1)^{k-\beta} \binom{\frac{ \nu}{\beta     }+m }{m}  \binom{\frac{ \nu}{\beta     }}{k-m-\beta}}{\frac{\nu}{\beta     } +m+\beta+\theta}
 = 
   \fracc{\Gamma(k+\theta)\Gamma(\beta+\frac{\nu}{\beta}+\theta)}{\Gamma(\beta+\theta)\Gamma(k+1+\frac{\nu}{\beta}+\theta)} 
}{iden_app_1}

The proof of this identity is omitted for space limitations. Using this to perform the summation in~\eqref{F_inv_1}, we obtain
\all{
F(z )   \xrightarrow{\mathcal{Z}^{-1} }
  \fracc{1      }{\beta     }  
   \fracc{\Gamma(k+\theta)\Gamma(\beta+\frac{\nu}{\beta}+\theta)}{\Gamma(\beta+\theta)\Gamma(k+1+\frac{\nu}{\beta}+\theta)} 
.}{F_inv}

This yields the inverse transform of the first term on the right hand side of~\eqref{psi_9}. For the second and third terms, we first ask: if the inverse transform of some function $F(z)$ is known, and is given by, say, $f_k$, then what is the inverse transform of $z^{\theta} \left(\frac{z-c}{1-c}\right)^{-\theta} F\left(\frac{z-c}{1-c}\right)$? We have: 
\all{
&z^{\theta} \left(\frac{z-c}{1-c}\right)^{-\theta} F\left(\frac{z-c}{1-c}\right) = 
(1-c)^{\theta} (z-c)^{-\theta} z^{\theta}\DD \sum_{r} f_r  \left(\frac{z-c}{1-c}\right)^{-r}
\nonumber \\ &
(1-c)^{\theta}  z^{\theta}\DD \sum_{r} \fracc{ f_r  (1-c)^r }{ ( z-c )^{r+\theta}}
=(1-c)^{\theta}  \DD \sum_{r} \fracc{ f_r  z^{-r}(1-c)^r }{ ( 1-c z^{-1} )^{r+\theta}}
\nonumber \\ &
=(1-c)^{\theta}  \DD \sum_{r} \sum_j   f_r  z^{-r}(1-c)^r  c^j   \binom{r+\theta+j-1}{j} z^{-j}
\nonumber \\ &
=(1-c)^{\theta}  \DD \sum_{k}  \Bigg[ \sum_r   f_r  (1-c)^r  c^{k-r}  \binom{k+\theta-1 }{k-r } \Bigg]z^{-k}
.}{steps}
So the inverse transform of $z^{\theta} \left(\frac{z-c}{1-c}\right)^{-\theta} F\left(\frac{z-c}{1-c}\right)$ is given by
\all{
 z^{\theta} \left(\frac{z-c}{1-c}\right)^{-\theta} F\left(\frac{z-c}{1-c}\right)
\Longrightarrow 
c^k (1-c)^{\theta} \sum_{r }  f_r \left(\frac{1-c}{c}\right)^r \binom{k+\theta-1}{r+\theta-1}
.}{inv_big}
Using this result, we can take the inverse transform of the other two terms on the right hand side of~\eqref{psi_9}. We obtain

\all{
&N_k(t)
= 
(1-c)^{\theta} c^k  \DD \sum_{r=1} ^{k} N(0,r,\theta)  \left( \fracc{1-c}{c} \right)^r \CC{k+\theta-1}{r+\theta-1}
\nonumber \\ &
 + \DD  
 \fracc{  \bigg[\lambda+ (2\beta +\theta)   t\bigg]   }{       \beta  }  
 \fracc{(k+\theta-1)!}{(\beta+\theta-1)!}  \fracc{\Gamma \left( \beta+ 2+\frac{\theta}{\beta} +\theta  \right) }
{\Gamma \left( k+3+\frac{\theta}{\beta}  + \theta \right) }  u(k-\beta)
\nonumber \\
&
\resizebox{.98 \linewidth}{!}{$
-  \fracc{   \lambda   (1-c)^{\theta} c^k  }{      \beta  }  
  \Gamma \left( \beta+ 2+\frac{\theta}{\beta} +\theta  \right)
\fracc{\Gamma(k+\theta) }{ \Gamma(\beta+\theta)}
\DD \sum_{r=\beta} ^{k}
 \displaystyle \frac{ \left( \fracc{1-c}{c} \right)^r  }{(k-r)! \Gamma \left( m+3+\frac{\theta}{\beta}  + \theta \right)}
$}.
}{N_ultimate_app}



\begin{thebibliography}{100}

\bibitem{BA_1}
  Barab{\'a}si , A.L.,   Albert,R. :  Emergence of scaling in random networks, Science  286, 509--512 (1999).
%
%
%
%
%
%
%
%
%
%
%
%

\bibitem{barab_link_growth_directed}
 Albert, R.,   Barab{\'a}si,A.L.,  : Topology of Evolving Networks: Local
  Events and Universality,  Phys. Rev. Lett. 85, 5234 (2000).

\bibitem{redner3}
Krapivsky, P. L., Rodgers, G. J.,   Redner, S.:Degree distributions of growing networks. Phy. Rev.  Lett. 86, 5401 (2001).
%

\bibitem{klemm_1}
 Klemm , K., Eguiluz,V.M. :Highly Clustered Scale-free Networks, Phys, Rev. E 65,  036123 (2002).

\bibitem{age_dorog_1}
  Dorogovtsev,S.,    Mendes,J.F.F. :Evolution of Networks with Aging of
  Sites,  Physical Review E 62, 1842 (2000).
%
%

\bibitem{newman_del_1}
 Moore, C.,  Ghoshal, G.,   Newman, M.E.J.:  Exact Solutions for Models of
  Evolving Networks with Addition and Deletion of Nodes,  Phys.l
  Rev. E 74,  036121 (2006).

\bibitem{sarshar}
 Sarshar , N.,   Roychowdhury,  V.: Scale-free and Stable Structures in Complex
  Ad Hoc Networks, Phys. Rev. E 69, 026101 (2004).
%
%

\bibitem{dorog_accel_1}
  Dorogovtsev , S.N.,   Mendes, J.F.F:  Effect of the Accelerating Growth of
  Communications Networks on their Structure, Phys. Rev. E 63,  025101 (2001).
 
%
%
%
%
%
%

\bibitem{kumar_copy}
 Kumar, R. ,  Raghavan, P.,  Rajagopalan, S., Sivakumar, D., Tomkins, A.,    Upfal, E.: 
   Stochastic Models for the Web Graph, Proc. 41st Symp.  Found.  of Comp.
  Sci.,  57--65 (2000). 

\bibitem{redner2}
Krapivsky, P. L., Redner, S., Leyvraz, F. : Connectivity of growing random networks.  Phys. Rev. Lett. 85,  4629 (2000). 



\bibitem{Bian}
 Bianconi , G.,    Barab{\'a}si, A.L.: Competition and multiscaling in evolving
  networks, Eur. Phys. Lett.  54, 436 (2001).
%

\bibitem{fitness_1}
 Smolyarenko, I., Hoppe, K.,  Rodgers, G.:   Network Growth Model with
  Intrinsic Vertex Fitness,  Phy. Rev. E 88,  012805 ( 2013).


\bibitem{pure_fit}
 Ghadge, S. , Killingback, T., Sundaram, B.,   Tran, D.A.:  A statistical
  construction of power-law networks,  Int. J.   Parallel
  Emergent and Distributed Systems 25, 223--235 (2010).

\bibitem{PDE}
 Zwillinger, D:   Handbook of Differential Equations,  ch. 2.
\newblock San Diego, CA: Academic Press, 1998.

\bibitem{ME_PRE}
 Fotouhi , B. ,  Rabbat, M.G.: Network growth with arbitrary initial
  conditions: Degree dynamics for uniform and preferential attachment, Phys. Rev. E 88,  062801 (2013).

\bibitem{dorog_nk}
Dorogovtsev, S. N., Mendes, J. F. F.,   Samukhin, A. N. : Structure of growing networks with preferential linking., Phys. Rev. Lett. 85,  4633 (2000).

\bibitem{redner1}
Krapivsky, P. L.,   Redner, S.: Organization of growing random networks, Phys.  Rev.  E 63, 066123  (2001).
%

\bibitem{Bol}
 Bollob{\'a}s, B., Riordan, O.,  Spencer, J., Tusnady, G.: The degree sequence
  of a scale-free random graph process,  Rand.  Struct.  and
  Alg. 18, 279--290 (2001).
%
%
\bibitem{chung}
Chung, F. R.,   Lu, L. : Complex graphs and networks :  CBMS Regional Conference Series in Mathematics 107,
 \newblock  American Mathematical Society (2006).

\end{thebibliography}
\end{document}